\documentclass[12pt]{article}
\usepackage{pdproc} 

\begin{document}

\renewcommand{\thefootnote}{\alph{footnote}}
  
\title{SOME USES OF NEUTRINO TELESCOPES\footnote{Invited talk presented
at Neutrino Telescopes Workshop, Venice, March 6-9, 2001; to be published
in the proceedings.}} 

\author{ SANDIP PAKVASA}

\address{ Department of Physics and Astronomy,
          University of Hawaii,
          2505 Correa Road\\
          Honolulu, HI  96822 \ USA\\
    {\rm     E-mail: pakvasa@phys.hawaii.edu} \\
    {\rm UH-511-985-01} \\
    {\rm  May 2001}}


%
%
\abstract{We imagine that large neutrino telescopes will be built and
  that distant neutrino sources of high energies and fluxes exist.  Some
  possible, if difficult, uses to which they might be put  are
  described; including (i) detecting neutrino mass-mixing possibly upto
  $\delta m^2$ of $10^{-16}  eV^2$; (ii) measuring neutrino cross-sections
   at PeV energies; (iii) detecting relic neutrinos; 
 (iv) doing cosmology with neutrinos and 
  (v) SETI with neutrinos.}
   
\normalsize\baselineskip=15pt
\section{Introduction}
  I make two optimistic assumptions.  The first one is that distant
  neutrino sources (e.g. AGN's and GRB's) exist; and furthermore with
  detectable fluxes at high energies (upto and beyond PeV).  The second
  one is that in the not too far future, very large volume, well
  instrumented detectors of sizes of order of KM3 and beyond will exist
  and be operating; and furthermore will have (a) reasonably good energy
  resolution, (b) good angular resolution ($\sim 1^0?)$ and (c) low
 energy threshold ($\sim 50  \ GeV?)$.  The ICECUBE discussed by Francis
  Halzen here is one example\cite{halzen}.

 If these assumptions are valid, then there are a number of uses these 
detectors can be put to.  By measuring the flavor mix of the neutrinos 
from a known source, mixing parameters can be determined for $\delta m^2$ 
as small as $10^{-16} eV^2$.  By measuring the attenuation of neutrinos 
in the earth, neutrino cross-sections at energies of PeV can be
determined.  
By confirming that the highest energy cosmic rays come from neutrinos
producing on-shell Z's, indirect evidence for relic neutrinos and their mass can be found.   
Measurements of flavor conversion probability (as a function of L/E) 
and of pulse spreading and separation can be start of neutrino cosmology 
(e.g. measuring red-shift in neutrinos as well $q_0$ and H).  Finally, a 
detection of a few neutrino events at energy of $M_Z/2$ could be the first 
sighting of an advanced extra-terrestrial civilization.
\section{Neutrinos from Active Galactic Nuclei}

	For AGN's the expectations are that they emit high energy $\nu's$;
the total flux probably overtakes atmospheric $\nu$-flux by $E_\nu \sim O 
(TeV)$ and the
most likely flavor mix is $\nu_\mu: \nu_e : \nu_\tau \approx 2:1:0.$

\subsection{$\nu_\tau$ Signature}
For a $\nu_\tau$ of energy above 2 PeV there is a characteristic ``double
bang'' signature\cite{learned}.   When $\nu_\tau$ interacts via charged current there is a
hadronic shower (of energy $E_1)$ with about $10^{11}$ photons emitted; then
the
$\tau$
travels about 90m (for $E_\tau \sim 1.8 PeV)$ and when it decays (either to
e's or hadrons with 80 \% probability) there is again a cascade (of energy
$E_2)$
with $2.10^{11}$ photons emitted in Cerenkov light.  The $\tau$ track is
minimum
ionising and may emit $10^6-10^7$ photons; even if it is not resolvable, one
can connect the two showers by speed of light and reconstruct the event.

The backgrounds (after appropriate cuts) are very small.  Hence such
``double bang'' events represent either $\nu_\mu \rightarrow \nu_e$ (or
$\nu_e \rightarrow \nu_\tau)$ oscillations or $\nu_\tau$-emission at the
source and in any case are extremely interesting.  For signal events due to
$\nu_\tau$, one expects $E_2/E_1 > 2$ on the average, and hence a cut of
$E_2/E_1 > 1$
removes many backgrounds; another cut
on the distance D between the two bangs of $D > 50m$ eliminates most of the
punch-thru backgrounds\cite{these}.

\subsection{Expected Flavor Mixes}
Most models of $\nu$-emission in AGN's correspond to tenuous beam dumps with
little absorption and $\nu's$ come from $\pi$ (and K) decay.  Frequently
$\gamma p \rightarrow  \Delta$ is a dominant process.  In these scenarios we
expect at production
\begin{eqnarray}
\nu_\mu: \ \nu_e: \ \nu_\tau \approx 2 : 1 : 0
\end{eqnarray}

For example, in the Protheroe-Szabo model\cite{protheroe}, they find $\nu_\mu :\nu_e
\approx 1.75:1$ and 10\% of $\nu's$ come from pp interactions.  Some
fraction of pp collisions will contribute to prompt $\nu's$ (including
$\nu_\tau's)$ via production of c and b. In the prompt $\nu's$ the flavor mix
is
\begin{eqnarray}
\nu_\mu: \nu_e:  \nu_\tau = 1 : 1 : p
\end{eqnarray}
where $p$ can be crudely estimated to be about $0.07$ to 0.1. Since the
prompt $\nu's$ themselves are expected to be only 10\% of total the modified
flavor mix is
\begin{eqnarray}
\nu_\mu:  \nu_e : \nu_\tau \approx 1: 0.6 : 0.01
\end{eqnarray}
and contains less than 1\% of $\nu_\tau's$.

\subsection{Rates}

To estimate event rates we make the following assumptions:
(i) assume the fluxes of Protheroe-Szabo model;
(ii) integrate over all AGN's;
(iii) assume an initial flavor mix of $\nu_\mu : \nu_e : \nu_\tau \approx
2:1:0$; and
(iv) a KM3 water or ice \^{c} detector with 100 \% detection efficiency\cite{we}.
Then we expect 1000 $\nu_\tau$ ``double bang'' events assuming maximal
$\nu-\mu_\tau$ mixing, 1000 $\nu_\mu$ events 
and about 1800 showering events $(\nu_e CC$ and $\nu_\alpha$ NC) per year.  With the
new upper limit from AMANDA presented here\cite{halzen}, this becomes
the upper bound.

\subsection{Flavor Mix on Arrival}

The neutrino flavor mix can be ``easily'' determined from the event
classification of the data.  The double bang events determine $\nu_\tau +
\bar{\nu}_\tau$ flux; the upcoming muons determine $\nu_\mu + \bar{\nu}_\mu$
flux; the cascade events (single bang) determine a
combination of $(\nu_e + \bar{\nu}_e), (\nu_\mu + \bar{\nu}_\mu)$ and
$(\nu_\tau
+
\bar{\nu}_\tau)$ fluxes; and Glashow Resonance (W) events determine
$\bar{\nu}_e$ flux (at $E_\nu = 6.4 PeV)$.

\subsection{Backgrounds}

We have considered several possible sources of backgrounds which fake
double bang signatures. The most serious appears to be a $\nu_\mu
\rightarrow \mu$ charged current event where the $\mu$ travels about 100 m
without much radiation and then deposits the bulk of its energy in a
catastrophic bremstrahlung.  This would have all the characteristics of a
genuine $\nu_\tau$ event.  We estimate the fraction of such events to be
about $(m_e/m_\mu)^2 (100 m/R_\mu)(\Delta E/E) \sim 3.10^{-3}$ and
seems reassuringly small.

At the hadronic vertex, the sources of background are:  (i) $\nu_e + N
\rightarrow e + D_s$ produced diffractively with $D_s \rightarrow \tau \nu$;
and $E_2/ E_1$ can be of 0(1) to fake the $\nu_\tau$ signal provided $D_s$
decays quickly.  The rate is expected to be of order $3.10^{-4}$ of cc events;
(ii) $\nu_{\alpha} + N \rightarrow \nu_\alpha + D_s/B$  again with $D_s$ or
B decaying into $\tau$ within 10m and $\tau$ traveling 100 m.  In these
events we expect $E_2/E_1 < 1$ and again the rate is small of order $\sim
10^{-3}$. Other backgrounds such as coincident downgoing $\mu's$ showering is
expected to be small.  Hence, that after the cuts such as $E_2/E_1 > 1
$ and $D > 50 m,$ the backgrounds are rather small.

We conclude that given AGN $\nu$-sources, it is possible to see $\nu_\tau
\rightarrow \tau$ events in a KM3 array unambigously.

\subsection{Sensitivity to Oscillations}

The sensitivity to oscillation parameters depends on several factors.  
If individual AGN's can be identified in $\nu_\tau's$ (say upto 100 Mpc or
more) then 
$\delta m^2$ upto $\geq 10^{-16} eV^2$ and mixing angles upto $sin^2 2
\theta \stackrel{\sim}{>} 0.05$ can be probed\cite{reno}.  On the other hand, if the 
current indications from atmospheric neutrino results are established as 
due to flavor oscillations, then the oscillating term in the conversion formula:
\begin{equation}
P_{\alpha \beta} = sin^2 2 \theta sin^2
\left ( \frac{\delta m^2 L}{4E} \right )
\end{equation}
averages to 1/2 and one can only confirm the expected value of mixing

To proceed further let us assume:  (i) initial fluxes are $\nu_\mu : \nu_e :
\nu_\tau \approx 2:1:0$; (ii) \# $\nu$ = \# $\bar{\nu}$ (although this is not
essential); (iii) all $\delta m^2 >> 10^{-16} eV^2,$
i.e.
$<sin^2~(\delta m^2 L/4
E) > \approx 1/2$; (iv) matter effects negligible at production (e.g. $N_{e-}
= N_{e +})$ and no significant matter effects en-route (this is valid for
$\delta m^2$ of current interest $\sim 10^{-2} - 10^{-6} eV^2)$\cite{see};(v)
atmospheric $\nu$-anomaly caused by $\nu_\mu - \nu_\tau$ oscillations with
$\delta m^2 \sim 5.10^{-3} eV^2$ and $\sin^2 2 \theta \geq 0.6$.  In this case we
expect
$\nu_\mu : \nu_e : \nu_\tau \approx 1: 1: 1$ at earth.

If should be stressed that this result viz. $\nu_\mu: \nu_e: \nu_\tau \cong$
1:1:1 depends crucially on the large $\nu_\mu - \nu_\tau$ mixing and is
relatively 
insensitive to the mixing of $\nu_e$\cite{this}. For example, the wide variety of
neutrino 
mixing matrices currently under consideration:  (i) Bi-maximal mixing,
(ii)Tri-maximal mixing, (iii) SMA 
(small angle MSW), (iv) Fritzsch-Xing mixing; all lead to the same result
as long as the initial flavor mix is $\nu_\mu: \nu_e: \nu_\tau = 2:1:\epsilon$.
Of course, if a source emits a universal flavor mix i.e  $\nu_\mu:
\nu_e: 
\nu_\tau = 1:1:1,$ it remains unchanged by oscillations.

Extra bonuses from observing the double-bang events are (i) the use of
the 
zenith angle distribution to measure $\nu_\tau$ cross section via
attenuation and 
(ii) use of enormous light collection and good timing to get good vertex
resolution 
and determine $\nu_\tau$ direction to within a degree or so.  Proposals
to 
account for the highest energy cosmic rays include some\cite{domokos} in which the
neutrino 
cross-sections are enhanced at very high energies. Because of
unitarity  
constraints, in the 2-20 PeV range they can increase by almost an order of
magnitude.  
Such scenarios can be possibly probed.

\section{Detecting Relic Neutrinos}

In the standard hot Big Bang Model\cite{seefor}, the effective temperature today of
relic neutrinos 
is $1.9^0K$; the number density per flavor is 110/cc (adding $\nu's$ and 
$\bar{\nu's})$; the average momentum is $5.2.10^{-4}$ eV/c.  
The current
density is 
$\sim 3.10^{12} cm^{-2} s^{-1}$ for massless neutrinos and $5.10^9 cm^{-2}
s^{-1}$for a neutrino mass of  $5.10^9 cm^{-2}s^{-1}$.
The $\nu_\alpha$-scattering cross section (at very low energies) for
Dirac neutrinos for allowed transitions  goes as
\begin{equation}
\sigma_\alpha    \sim   \frac{a^2_\alpha \ G_F^2 \ m_\nu^2}{\pi}
\end{equation}
where $a_\alpha\ = (3Z-A)$ for $\alpha=e$ and $a_\alpha\ = (A-Z)$
for $\alpha=\mu$ or $\tau$.
Many early proposals to detect relic neutrinos by reflection or coherent
effects turned out to be incorrect.  There are three methods which 
some day may prove to be practical.

The first is a 1975 proposal due to Stodolsky\cite{stodolsky}.  The idea needs 
neutrino asymmetry i.e. excess of $\nu$ (or $\bar \nu)$  over
$\bar{\nu}$ (or $\nu)$ in order to
work.  Then a polarized electron moving in a background of CMB neutrinos 
can change its polarization due to the axial vector parity violating interaction.  
The effective interaction goes as
\begin{equation}
H_{eff} \sim \frac{2 G_F}{\sqrt{2}} \ \underline{\sigma}_e.\underline{v} \
n_\nu
\end{equation}
With $v \sim 300 \ km s^{-1}$ and $n_\nu \sim 10^7/cc$ this correspond to an
energy of about $10^{-33} eV$ and 
leads to a rotation of the polarization of about 0.02" in a year.  
Can such small spin rotations can be detected?  Certainly not at 
present, but technology may someday allow this.

The second method is one suggested by Zeldovich and 
collaborators~\cite{shvartsman}.  The
idea is to take advantage of momentum transfer in neutrino-nucleus
scattering.  Consider an object made up of small spheres of radius $a
\approx \lambda$ (neutrino wavelength) packed loosely with pore sizes
also of the same size.  (to avoid destructive interference).  If the
number of atoms in the target is $N_A$, then the effective coherent
cross-section is
\begin{equation}
\sigma   =   \sigma_\alpha \ N_A^2
\end{equation}
where $\sigma_\alpha$ is as given in Eq. (10).  Assuming total reflection,
momentum transfer is
\begin{equation}
\Delta p \cong 2m_\nu \ v_\nu
\end{equation}
and the force $f=j_\nu \sigma \Delta p$ is given by
\begin{equation}
f=2 n_\nu \sigma_\alpha \ N_A^2   m_\nu \ v^2_\nu
\end{equation}
The most optimistic estimates are obtained by assuming some clustering
$(n_\nu \sim 10^7/cc), m_\nu \sim 0 (eV), v_\nu \sim 10^7 cms^{-1}, \rho \sim
20 gm/cc$; leading to
\begin{equation}
a= \frac{f}{m} = \  \frac{f}{N_A m_N}  \ \sim 10^{-23}  (a_\alpha/A)^2   \
cm.s^{-2}
\end{equation}
Such accelerations are at least ten orders of magnitude removed from
current sensibility and possible detection remains far in future~\cite{several}.
Incidentally, this is based on having Dirac neutrinos; for Majorana
neutrinos, one would need spin alignment in a macroscopic sample.
   
   The third possibility is the one proposed by Weiler in
1984~\cite{weiler}.  The basic
idea is as follows.  If neutrinos have masses in the eV range and there
are sources of very high energy neutrinos at large distances, then the
H.E. $\nu$ can annihilate on the C.B.R. $\bar{\nu}$ and make a $Z^0$
on-shell at resonance creating an absorption dip in the neutrino
spectrum.  The threshold for $Z$ production would be at $E \sim
m_Z^2/2m_\nu$ which is about $4.10^{21}$ eV for an eV neutrino mass.  This seemed like an
unlikely possibility, since it required large neutrino fluxes at very
high energies to see the neutrino spectrum and then the absorption
dip.  But all this changed dramatically recently with the clear signal
of cosmic rays beyond the GZK cut-off~\cite{greisen}.  The GZK cut-off is the energy at
which cosmic ray protons pass the threshold for pion production off the
CMB photons.  This is at an energy $E \sim m_\pi mp/E_\gamma \sim
6.10^{19}$ eV.  Above this energy, the mean free path of protons is less
than 10 Mpc and hence these protons have to be ``local''.  The flux
should then decrease dramatically since we believe the cosmic rays are
not produced locally.  Recently, what used to be hints of the cosmic ray
signal extending beyond this cut-off, has become a clear signal~\cite{takeda}.  The
events are most likely due to primary protons.  Then an explanation is
called for.  One intriguing proposal~\cite{weiler1} is that these events are nothing
but a signal for the Z's produced by the $\nu \bar{\nu} \rightarrow Z$
process with the protons coming from the subsequent Z decay!  Of
course, the original problem of needing sources of high energy neutrinos
remains.  If this explanation is valid, we have already seen (indirect)
evidence for the existence of relic neutrinos.  In principle, this
proposal can be tested:  (i)  the events should point back  at the
neutrino sources; (ii) there is an eventual cut-off when the energy
reaches the threshold energy for $Z$ production, $E \sim 4.10^{21}
\left ( \frac{eV}{m_\nu} \right)$ eV; (iii) $\gamma/p$ ratio should be large
near threshold and (iv) the large $\nu$-flux should be eventually seen
directly in large $\nu$-telescopes. There is the bonus that the cut-off
energy also measures the mass of the relic neutrinos! Thus neutrino telescopes
can give existence proof of relic neutrinos as well as measure their
mass.

\section{Cosmology with neutrinos}

We know from supernova studies that there are several effects of
neutrino masses and mixings on the observation of neutrino bursts.  A
pulse spreads in time due to dispersion of velocities (from non-zero
mass); a pulse separates into several pulses due to a neutrino of a
given flavor being a mixture of different mass eigenstates  and the original
flavor composition can change due to mixing and oscillations. One can apply these
considerations to neutrino pulses from sources which are at cosmological
distances.  Then the effects come to depend on cosmological parameters. 

For example, the time difference~\cite{stodolsky1} between two mass eigen-states which
left at the same time is given by 
\begin{equation}
\Delta t \approx z/H \left [1-\frac{(3+q_0)}{2} z .... \right ]  \frac{1}{2} 
\left [ \frac{m_1^2}{E_1^2} - \frac{m_2^2}{E_2^2} \right ]
\end{equation}
where $E_i$ are the energies observed at earth and $z,H$ and $q$ have the
usual meanings.  The spreading of a pulse of a given mass neutrino is
given by~\cite{stodolsky1}
\begin{equation}
\Delta t \approx z/H \left [1-\frac{(3+q_0)}{2}  z ..... \right ]  \frac{1}{2} m^2 
\left \{ \frac{1}{E_1^2} - \frac{1}{E_2^2} \right \}
\end{equation}

Finally, the conversion probability for an emitted flavor $\alpha$ to
become $\beta$ at detection is given by

\begin{equation}
P_{\alpha \beta} = sin^2 2 \theta \ sin^2 \phi/2
\end{equation}
where the phase $\phi$ is~\cite{weiler2}
\begin{equation}
\phi \cong z/H \left [ 1-\frac{(3+q_0)}{2}  z .... \right ]  \frac{\delta m^2}{2E}
\end{equation}

The basic flight time factors are rather small, for eV neutrino masses
and GeV energies, $\Delta t \sim$ 50 milliseconds at 1000 MPc.  
These time spreads and separation may be shorter than the times involved
in the production
process thus making them difficult to observe.  
On the other hand if the current suggestions for O(KeV) masses\cite{giudice} for
$\nu_\mu$ 
and $\nu_\tau$ to act as warm dark matter\cite{bode} turn out to be valid, then
time 
delays are much larger (e.g. O(sec) at distances of Mpc).

As for flavor conversion, emitted $\nu_\mu's$ can get converted into
$\nu_\tau's$ and thus produce a significant incoming flux of
$\nu_\tau's$ (which is essentially absent initially in most neutrino
production scenarios).  With the flavor mix of the incoming beam
determined as discussed above, $P_{\alpha
\beta}$ and hence the phase $\phi$ can be deduced by comparing to
expected initial relative fluxes.  Provided the phase $\phi/2$ is not too large
(and $\sin^2 \phi/2$ does not average to 1/2) one has sensitivity to the
parameters $z,q_0,$ and $H$.

With such measurements of $\Delta t$ and $\phi$, one can potentially
measure these cosmological parameters.  This would be the
first time that the red-shift or  other cosmological
parameters are measured for anything other than light.  There is another advantage of
using neutrinos.  This is the fact that the initial flavor mixing only
depends on microphysics and so the comparison is free from problems such
as evolution or worries about standard candles etc.

\section{SETI with Neutrino}

 Yes, I do mean\cite{learned1} search for extra-terrestrial intelligence. 
There is a school of thought that holds that this search in futile, 
pointless,  and bound to yield null results.  However, as has been 
pointed out, absence of evidence is not necessarily evidence of absence.

 Imagine that an advanced civilization exists in the galaxy, with many
outposts. It will need to maintain time standards over a long base line.  In turn, 
this will require (i) stable clocks of high precision, (ii) fast
processes for transmitting and receiving time markers and (iii)  a form of
radiation which will faithfully carry timing data over long distances.

 The need for clock synchronization stems from the fact that standard
clocks have to exchange timing data to remain synchronized.  This is in order
to correct for general relativistic effects which depend on positions and
motions of nearby massive objects. Furthermore, the presence of chaos in many 
body systems means that such corrections cannot be calculated indefinitely 
from initial data alone, so that the synchronization has to be done repeatedly.

 The requirements of a mobile, spread out civilization would suggest the
use of isotropic synchronization signals.  Other arguments suggest the same
thing. Hence, even though it raises the required energy budget, this is the 
most likely scenario.

 The fastest known process is the $Z^0$ decay with a lifetime of $2.5\times
10^{-25}s$. It also produces neutrinos of  45.6 GeV, satisfying the
requirements of radiation which can carry information intact thru many obstacles.

 If an advanced civilization is using this process to send timing
signals, a neutrino telescope can detect some neutrino events at  the
energy of 45.6 GeV. If the source is a few kpc from us, then a KM3 water/ice
\^{c} detector will detect a few events per year (all flavors in equal numbers).

 The ETI would have to overcome many technical problems to implement such
a scheme.  We have addressed some of them elsewhere\cite{learned1}.  The power
requirements 
to give a few events per year in a KM3 detector at a distance of few kpc
are 
huge; approximately the solar luminosity $ \sim 10^{45} eV/sec.$  This,
of course, 
is {\it their} problem and we have to imagine that they have solved it.  Is it
possible 
that a technology radiating such huge amounts of power within a few kpc
has escaped our detection?
  We speculate that this would correspond to a 
"Dyson shell." Dyson had suggested that if an advanced civilization 
surrounds a star with a shell of material and uses heat engines to
extract power, 
then the system would appear as  an infra-red source.  Since the IRAS data 
include over 50,000 IR sources, some of these indeed could well be "Dyson shells".  

 These synchronization neutrino signals at E =  $M_Z/2$ are extremely distinctive 
in that they are not expected to occur naturally and are therefore
unlikely to be mistaken for anything else.  In view of the spectacular 
nature of the 
timing signal and the enormous implications of its detection, we believe
it is surely worth keeping watch for it.

\section{Conclusion}

Neutrino telescopes are "Field of Dreams"!  If we build them; the neutrinos, they will come!

\section{Acknowledgement}

I thank all my collaborators on whose work this is based:  John Learned, 
Walter Simmons, Tom Weiler and Xerxes Tata; and Milla Baldo-Ceolin for 
another wonderful experience in Venice.  This work was supported in part 
by U.S.D.O.E. under Grant DE-FG 03-94ER40833.

\end{document}